\documentclass[10pt]{article}

\usepackage[margin=0.78in]{geometry}
\usepackage[T1]{fontenc}
\usepackage[utf8]{inputenc}
\usepackage{lmodern}
\usepackage{amsmath,amssymb}
\usepackage{booktabs,longtable,array}
\usepackage{microtype}
\usepackage{xcolor}
\usepackage{xurl}
\usepackage[hidelinks]{hyperref}

\newcommand{\concat}{\mathbin\Vert}

\newcommand{\SHA}{\operatorname{SHA256}}
\newcommand{\evidence}[1]{\textsuperscript{\scriptsize\textsf{[#1]}}}
\newcommand{\currentv}{\textsc{current V2}}
\newcommand{\historicalv}{\textsc{historical V1}}
\newcommand{\boundary}{\textsc{boundary}}

\newcommand{\MutMemEvidenceSourceCommit}{\nolinkurl{eb0166fd45964ebb47ca053bcfeb0ab3ddc68999}}
\newcommand{\MutMemPublicationEvidenceRoot}{\nolinkurl{99bed29dcf0cc5c2f597f17b69be8d6dc990bae4ae5dd30838a1d1f34d92a76b}}
\newcommand{\MutMemClaimMapRoot}{\nolinkurl{4d8d23442e9ed311e33ddd9ce9a79e878642eba93329c460e203128e54997527}}
\newcommand{\MutMemIndependentVerificationRoot}{\nolinkurl{f38268c86f2291971ac6ab75ded944d9a7e7eb6fa2809a5367d69dacd450ffe0}}

\newcommand{\MutMemRepositoryURL}{\url{https://github.com/wallidsaydi-creator/HOM-AIMOS}}
\newcommand{\MutMemReleaseURL}{\url{https://github.com/wallidsaydi-creator/HOM-AIMOS/releases/tag/v1.0.0}}
\newcommand{\MutMemReleaseTag}{v1.0.0}
\newcommand{\MutMemReleaseCommit}{\nolinkurl{fcda26d1e05c729185c09dd11446db2f79c14a0d}}
\newcommand{\MutMemSourceManifestRoot}{\nolinkurl{f0cdb75763c70ce83bfd1e5f6c956e7502b65a812c3a43f311f8039ba0c3b7f1}}
\newcommand{\MutMemVOneArxiv}{arXiv:2608.02843}
\newcommand{\MutMemProtocolSchemaCount}{18}
\newcommand{\MutMemRecallStructuralVectorCount}{39}
\newcommand{\MutMemMutationStructuralVectorCount}{15}
\newcommand{\MutMemFailureCodeCount}{37}
\newcommand{\MutMemCrossLanguageTerminalCount}{72}
\newcommand{\MutMemVerifierExitPassed}{10}
\newcommand{\MutMemVerifierExitTotal}{10}
\newcommand{\MutMemConformanceCount}{42}
\newcommand{\MutMemConformanceClassCount}{28}
\newcommand{\MutMemCanaryPopulation}{120}
\newcommand{\MutMemCanaryMarkedDetected}{60}
\newcommand{\MutMemCanaryMarkedTotal}{60}
\newcommand{\MutMemCanaryMarkedPercent}{100.0}
\newcommand{\MutMemCanaryMarkedWilsonLower}{93.98}
\newcommand{\MutMemCanaryCleanFlagged}{0}
\newcommand{\MutMemCanaryCleanTotal}{60}
\newcommand{\MutMemCanaryCleanPercent}{0.0}
\newcommand{\MutMemCanaryCleanWilsonUpper}{6.02}
\newcommand{\MutMemInstallerNodeVersion}{v26.8.1}
\newcommand{\MutMemInstallerGuideCount}{8}
\newcommand{\MutMemInstallerExperimentalCount}{0}
\newcommand{\MutMemInstallerClearanceMaximum}{10}
\newcommand{\MutMemLongMemCorrect}{459}
\newcommand{\MutMemLongMemTotal}{500}
\newcommand{\MutMemLongMemPercent}{91.80}
\newcommand{\MutMemLongMemWilsonLower}{89.06}
\newcommand{\MutMemLongMemWilsonUpper}{93.90}
\newcommand{\MutMemLocomoCorrect}{1472}
\newcommand{\MutMemLocomoTotal}{1986}
\newcommand{\MutMemLocomoPercent}{74.12}
\newcommand{\MutMemLocomoWilsonLower}{72.15}
\newcommand{\MutMemLocomoWilsonUpper}{76.00}
\newcommand{\MutMemLocomoFOne}{58.20}
\newcommand{\MutMemPoisonCleanAsrCount}{2}
\newcommand{\MutMemPoisonCleanAsrTotal}{100}
\newcommand{\MutMemPoisonAttackedAsrCount}{3}
\newcommand{\MutMemPoisonAttackedAsrTotal}{100}
\newcommand{\MutMemPoisonInducedCount}{1}
\newcommand{\MutMemPoisonInducedTotal}{98}
\newcommand{\MutMemPoisonInducedPercent}{1.02}
\newcommand{\MutMemPoisonRetrievalCount}{0}
\newcommand{\MutMemPoisonRetrievalTotal}{100}
\newcommand{\MutMemPoisonRetrievalWilsonUpper}{3.70}
\newcommand{\MutMemPoisonCleanAccuracyPercent}{72.0}
\newcommand{\MutMemPoisonAttackedAccuracyPercent}{71.0}
\newcommand{\MutMemPoisonLabelCount}{500}
\newcommand{\MutMemPoisonLabelTotal}{500}
\newcommand{\MutMemCleanAdverseCount}{4}
\newcommand{\MutMemCleanAdverseTotal}{18808}
\newcommand{\MutMemCleanAdversePercent}{0.0213}
\newcommand{\MutMemMutationTransitionCount}{20}
\newcommand{\MutMemMutationAuthorizationPassed}{7}
\newcommand{\MutMemMutationAuthorizationTotal}{7}
\newcommand{\MutMemMutationTamperPassed}{4}
\newcommand{\MutMemMutationTamperTotal}{4}
\newcommand{\MutMemMutationMedianMs}{4.865}
\newcommand{\MutMemMutationPNinetyFiveMs}{5.674}
\newcommand{\MutMemMutationMeanBytes}{966.35}
\newcommand{\MutMemHumanAgreementCount}{193}
\newcommand{\MutMemHumanAgreementTotal}{200}
\newcommand{\MutMemHumanAgreementPercent}{96.5}
\newcommand{\MutMemHumanKappa}{0.9108}
\newcommand{\MutMemStatisticalRateRows}{44}
\newcommand{\MutMemStatisticalInferenceRows}{28}
\newcommand{\MutMemEvidenceManifestOutputCount}{7}
\newcommand{\MutMemAblationAZeroRetrievalCount}{94}
\newcommand{\MutMemAblationAZeroRetrievalTotal}{100}
\newcommand{\MutMemAblationAZeroCleanPercent}{71.0}
\newcommand{\MutMemAblationAZeroAttackedPercent}{40.0}
\newcommand{\MutMemAblationAOneRetrievalCount}{0}
\newcommand{\MutMemAblationAOneRetrievalTotal}{100}
\newcommand{\MutMemAblationAOneCleanPercent}{68.0}
\newcommand{\MutMemAblationAOneAttackedPercent}{65.0}
\newcommand{\MutMemAblationATwoRetrievalCount}{0}
\newcommand{\MutMemAblationATwoRetrievalTotal}{100}
\newcommand{\MutMemAblationATwoCleanPercent}{68.0}
\newcommand{\MutMemAblationATwoAttackedPercent}{71.0}
\newcommand{\MutMemAblationAThreeRetrievalCount}{0}
\newcommand{\MutMemAblationAThreeRetrievalTotal}{100}
\newcommand{\MutMemAblationAThreeCleanPercent}{69.0}
\newcommand{\MutMemAblationAThreeAttackedPercent}{69.0}

\title{MutMem V2: Cryptographically Authorized Mutation in\\
Persistent Agent Memory\\[0.35em]
\large Portable Verification and Reproducible Evidence}
\author{Walid Saidi\\Independent researcher}
\date{Evidence-audited manuscript candidate, September 2026}

\begin{document}
\maketitle

\begin{center}
\fbox{\begin{minipage}{0.96\linewidth}
\small
\textbf{Reproducibility and release status.}
Repository: \MutMemRepositoryURL. Published predecessor: \MutMemVOneArxiv.
The evidence source commit is \MutMemEvidenceSourceCommit, and the closed
publication-evidence root is \MutMemPublicationEvidenceRoot.

The public release is \MutMemReleaseURL, immutable tag \MutMemReleaseTag,
commit \MutMemReleaseCommit, with source-manifest root
\MutMemSourceManifestRoot.

Offline verification is exactly \texttt{npm run verify}; deterministic evidence
regeneration is exactly \texttt{npm run evidence:regenerate}. For an ordinary
agent identifier stored in shell variable \texttt{AGENT\_ID} after installer
onboarding, the exact multi-line utility reproduction command is:
\begin{center}
\texttt{npm run reproduce -{}- -{}-installed-instance canonical \textbackslash}\\
\texttt{~~-{}-agent-id "\$AGENT\_ID" -{}-full -{}-benchmark both \textbackslash}\\
\texttt{~~-{}-protocol canonical-blind-v1}
\end{center}

Source code is AGPL-3.0-or-later. The repository license applies to
AIMOS-authored artifact files. Dataset text is not redistributed; downloaded
materials retain upstream licenses. Historical V1 utility used GPT-5.4 for
generation and GPT-5.6 Terra for judgment; the historical poisoning lane used
GPT-5.5 and GPT-5.6 Terra. Provider access is operator-authenticated, and no
credential or provider payload is released.

\textbf{Highest support level:} the protocol and evidence are independently
verifiable, and one clean installation is qualified; the empirical findings
have not been independently replicated. The immutable public tag, release
commit, GitHub Release, and source-manifest root are mutually bound. This
manuscript is publication-authorized for arXiv submission; the arXiv identifier
is necessarily recorded after submission acceptance.
\end{minipage}}
\end{center}

\begin{abstract}
MutMem V1, published as ``MutMem: Cryptographically Authorized Mutation in
Persistent Agent Memory,'' introduced retention-preserving, cryptographically
authorized mutation for persistent agent memory and reported utility,
mutation-integrity, and targeted-poisoning results. Its principal publication gap was not another
retrieval mechanism: it was the absence of a complete portable verification
contract and a reviewer-facing path from a clean installation to independently
checkable evidence. MutMem V2 closes that gap without creating a second memory
engine. It specifies exact canonical bytes, domain-separated object and bundle
commitments, mandatory recall-evidence membership and order, external trust
anchors, identity epochs, revocation state, authorization, request receipts,
ordered disclosure, and three mutation terminals. The current protocol contains
\MutMemProtocolSchemaCount\ versioned object schemas,
\MutMemRecallStructuralVectorCount\ recall predicate vectors,
\MutMemMutationStructuralVectorCount\ mutation vectors, and
\MutMemFailureCodeCount\ closed recall failure reasons. Independent Node and
Python implementations agree on verdict and primary reason for all
\MutMemCrossLanguageTerminalCount\ structural and cryptographic terminals; a
separate production-conformance corpus agrees on
\MutMemConformanceCount/\MutMemConformanceCount\ cases spanning
\MutMemConformanceClassCount\ required classes. A clean Node
\MutMemInstallerNodeVersion\ installation reaches first-boot, restart, and
scheduler readiness with no experimental memories. A separately scoped Canary
experiment completes \MutMemCanaryPopulation\ explicit-marker and clean units;
it is evidence of declared marker traversal only. Every public table is
regenerated from a self-hashed aggregate, and an independent verifier
reconstructs its statistics and claim boundaries. Historical V1 empirical
results remain historical rather than being relabeled as V2 reruns. MutMem V2
supports claims about portable integrity, authorization, traceability,
conformance, and reproducibility under stated assumptions. It does not establish
semantic truth, universal robustness, or independent replication.
\end{abstract}

\section{Introduction}

Persistent memory creates two separable research problems. The first is how an
agent may adapt retained evidence while preserving authorship, prior state, and
the reason for change. MutMem V1 addressed that problem with append-only outcome
evidence, bounded bidirectional weight transitions, signed provenance, and
ordered recall receipts \cite{mutmemv1}. The second problem is how a reviewer can
verify those claims without trusting the production database, its runtime, or a
prose description of the system. A self-consistent implementation is not yet a
portable protocol: its object boundaries, byte encodings, trust roots, failure
reasons, result cardinality, and terminal states must be explicit.

MutMem V2 addresses the second problem. MutMem denotes the portable protocol;
HOM-AIMOS is one producer and case study. The portable verifier does not select
memories, authorize requests, sign events, mutate storage, or operate a second
runtime. It consumes finite evidence objects and returns a deterministic terminal
verdict. \evidence{V2-PORTABLE-PROTOCOL}

The paper makes four contributions:

\begin{enumerate}
  \item It defines a versioned recall-disclosure envelope whose complete
  membership is $13+5r$ objects for $r$ disclosed results, with deterministic
  ordering, exact canonical bytes, and domain-separated SHA-256 commitments.
  \item It defines cross-object predicates for external trust, actor and
  Housekeeper epochs, revocation, effective authorization, signed request and
  receipt parity, native decision bindings, per-result provenance and occurrence
  evidence, Merkle order, and signed terminal receipts.
  \item It defines a portable mutation profile that preserves the native outcome
  schema and distinguishes an authorized transition, a signed no-op, and an
  occurrence observation.
  \item It supplies independent Node and Python verifiers, a production-derived
  conformance corpus, a clean installer qualification, deterministic publication
  regeneration, an attempt ledger, and a complete claim-to-evidence map.
\end{enumerate}

The contribution is deliberately narrower than ``secure memory.'' Cryptographic
integrity does not imply that remembered content is true. A fixed poisoning
experiment does not imply universal detection. Exact verifier parity does not
imply independent empirical replication. These are protocol boundaries, not
disclaimers added after observing results. \evidence{CONTENT-TRUTH}
\evidence{UNIVERSAL-DEFENSE}\evidence{INDEPENDENT-REPLICATION}

\section{Evidence identities and support levels}

We use three non-interchangeable evidence labels:

\begin{itemize}
  \item \currentv{} denotes evidence produced or reverified for the portable V2
  protocol, independent verifiers, conformance corpus, explicit-marker Canary
  lane, or clean installer.
  \item \historicalv{} denotes immutable empirical evidence released with
  MutMem V1. It may be cited as a historical comparator but is not a current
  V2 benchmark run.
  \item \boundary{} denotes an omitted or prohibited claim. Missing evidence is
  never converted into a zero, success, or defense.
\end{itemize}

The manuscript binds the claim map \MutMemClaimMapRoot\ and independent
verification \MutMemIndependentVerificationRoot. The claim map names the code,
protocol, artifact, verifier, denominator, and limitation for every quantitative
or security statement. The evidence package contains
\MutMemEvidenceManifestOutputCount\ deterministic files including its manifest;
the manuscript assets are generated separately from that root.

\section{System and threat model}

\subsection{Principals and boundary}

The production producer has an external master trust anchor, an ordinary actor
or the Housekeeper system principal, append-only identity epochs and revocation
state, a signed authorization projection, a canonical SAVE path, a canonical
RECALL path, and restricted database-local writers. The portable verifier is an
evidence consumer. It neither possesses producer credentials nor infers a trust
root from the bundle it is asked to verify.

For ordinary actors, the evidence bundle must bind the exact company, actor
identifier, identity epoch, certificate fingerprint, unrevoked state, effective
grant, requested clearance, data class, method, path, canonical request body,
nonce, and signing time. For the Housekeeper, the system-principal profile is
structurally distinct from a master-signed ordinary grant; signature-bearing
grant fields are forbidden in that profile. This separation prevents a system
principal from masquerading as an ordinary master-signed grant and prevents an
ordinary identity from inheriting Housekeeper authority.

\subsection{Adversary}

The verifier is designed to detect malformed canonical objects, object or bundle
substitution, wrong trust roots, identity or epoch substitution, stale or revoked
authority, request replay or route mismatch, result omission or reordering,
provenance and occurrence substitution, Merkle mismatch, terminal-receipt drift,
and mutation-topology inconsistency. The adversary may control transport and may
modify copied evidence. The adversary may not break SHA-256 collision resistance
or Ed25519 existential unforgeability, and the reviewer must obtain the expected
master fingerprint and release identity through an external channel.

An attacker who replaces both the evidence and every external trust anchor can
present a different self-consistent world. MutMem does not solve that anchor
distribution problem. It also does not prove availability, confidentiality,
semantic truth, model honesty, or protection against every form of memory
poisoning.

\section{Portable recall-disclosure protocol}

\subsection{Canonical objects}

Let $H=\SHA$, let $J(x)$ be the protocol's deterministic canonical-JSON bytes,
let $\operatorname{LP}_4(x)$ prepend the unsigned four-byte big-endian byte
length of UTF-8 string $x$, and let $\operatorname{BE}_k(x)$ encode integer $x$
in $k$ big-endian bytes. Every variable-length field is either canonical JSON or
explicitly length-framed. The object domain $D_o$ is the UTF-8 byte string
\texttt{hom.aimos.mutmem-portable-object/v2} followed by a zero byte. For kind
$k_j$, schema $s_j$, and body $b_j$, the object commitment is

\begin{equation}
o_j = H\!\left(
D_o \concat \operatorname{LP}_4(k_j)
\concat \operatorname{LP}_4(s_j)
\concat \operatorname{BE}_4(|J(b_j)|)
\concat J(b_j)
\right).
\label{eq:object}
\end{equation}

Bodies reject unsupported or non-finite numbers, unsafe integers, nesting beyond
the protocol limit, and bodies larger than the declared maximum. Object schemas
are exact, not advisory labels. These choices follow the prefix-free framing and
domain-separation discipline used in cryptographic protocol design
\cite{bonehshoup} and the deterministic JSON discipline standardized by JCS
\cite{rfc8785}.

\subsection{Complete membership and deterministic order}

Each recall bundle contains thirteen singleton objects: trust anchor; actor
identity and revocation; Housekeeper identity and revocation; effective grant;
request envelope; request receipt; content-state decision; epistemic decision;
final security closure; return projection; and native recall receipt. Each
disclosed result contributes five objects: memory state, provenance chain,
occurrence, epistemic projection, and receipt evidence. Therefore, for $r$
results,

\begin{equation}
N_{\mathrm{objects}}(r)=13+5r, \qquad 0\le r\le 200.
\label{eq:cardinality}
\end{equation}

Every singleton kind occurs exactly once. Every result ordinal from $0$ through
$r-1$ contains all five result kinds and one subject identifier shared across
the group. Objects are ordered first by the fixed singleton order and then by
result ordinal and fixed result-kind order. Duplicate, missing, mixed-subject,
or out-of-range objects fail before relational evaluation.

The ordered object descriptors are committed with the RFC 6962 construction
\cite{rfc6962}. If $x_i$ denotes canonical bytes for descriptor $i$,

\begin{align}
L_i &= H(\texttt{0x00}\concat x_i), \\
T(a,b) &= H(\texttt{0x01}\concat a\concat b),
\label{eq:merkle}
\end{align}

with an empty root $H(\epsilon)$ and recursive split at the largest power of two
strictly smaller than the current list length. Leaf/node domain separation
prevents a node encoding from being confused with a leaf encoding under the
hash assumption.

\subsection{Envelope commitment}

Let $D_b$ be \texttt{hom.aimos.mutmem-portable-evidence/v2} followed by a zero
byte, $u$ the bundle identifier, $c$ the company identifier, $f$ the externally
expected 32-byte master fingerprint, $r$ the result count, and $R$ the object
root. The bundle commitment is

\begin{equation}
B = H\!\left(D_b\concat \operatorname{LP}_4(u)
\concat \operatorname{LP}_4(c)\concat f
\concat \operatorname{BE}_8(r)\concat R\right).
\label{eq:bundle}
\end{equation}

The master fingerprint is an input to verification; accepting the fingerprint
solely because it appears inside the bundle would be circular trust. The
structural predicate evaluator therefore reports that cryptographic signatures
and external trust remain unverified. Independent verifiers must check them.

\subsection{Cross-object predicates}

The recall predicate graph enforces exact equality across five boundaries:

\begin{enumerate}
  \item \textbf{Authority:} trust anchor, actor epoch, certificate, revocation,
  Housekeeper epoch, authorization, and request time agree.
  \item \textbf{Admission:} signed method and path are \texttt{POST} and
  \texttt{/aimos/recall}; request-body, normalized-command, request-receipt,
  nonce, signature, and authorization mutation commitments agree.
  \item \textbf{Decision composition:} content-state roots, epistemic decision,
  final security closure, return path, projected identifiers, and content hashes
  agree, with no canonical-memory or retention mutation.
  \item \textbf{Per-result evidence:} memory, provenance, occurrence, epistemic,
  and receipt objects agree on subject, live-content hash, save and binding
  mutations, occurrence reference, ordinal, and selected membership.
  \item \textbf{Terminal receipt:} Merkle entries, root, result count, actor,
  request, authority, Housekeeper certificate, event content hash, predecessor,
  nonce, signing time, signature, and revocation evaluation agree.
\end{enumerate}

The failure vocabulary is closed: an undeclared failure reason is itself an
error. This makes verdict/reason parity testable across implementations instead
of permitting semantically similar but incomparable exceptions.

\section{Portable mutation profile}

MutMem V2 does not introduce a new cognitive update rule. It preserves the
native V1 mutation-outcome schema and makes its evidence portable. Let $D_\mu$
be \texttt{hom.aimos.mutmem-portable-mutation-evidence/v2} followed by a zero
byte, and let $M$ be the canonical body containing the recall commitment,
outcome evidence, recall receipt, outcome event, signed valence evidence,
terminal, and optional cognitive projection. The portable mutation commitment is

\begin{equation}
B_\mu = H(D_\mu\concat J(M)).
\label{eq:mutation-bundle}
\end{equation}

Three terminals are exhaustive:

\begin{itemize}
  \item \texttt{authorized\_transition}: a principal-state outcome binds a
  terminal REWEIGHT provenance node and a nontrivial cognitive projection;
  \item \texttt{signed\_noop}: the signed outcome is retained, but no projection
  is fabricated when the quantized weight is unchanged; and
  \item \texttt{occurrence\_observation}: occurrence-scoped evidence is retained
  with no weight authority.
\end{itemize}

For an authorized transition, the verifier reconstructs the V1 projection and
transition hashes from the memory identifier, integer milliscale old/new
weights, provenance mutation, predecessor projection, and company. It requires
the retained Ed25519 signature and exact signer epoch; Ed25519 itself is a
standard primitive, not a MutMem contribution \cite{rfc8032}.
\evidence{V1-MUTATION-INTEGRITY}

\section{Independent verification}

\subsection{Two implementations, one terminal language}

The production-language reference and independent Python implementation consume
the same public vectors but do not share verifier code. Four lanes cover recall
structure, mutation structure, recall cryptography, and mutation cryptography.
Across all \MutMemCrossLanguageTerminalCount\ vectors, both implementations
match the expected valid/invalid terminal and exact primary reason. This count is
not a benchmark accuracy denominator: it is a conformance denominator.
\evidence{V2-INDEPENDENT-VERIFIER}

The structural vector sets contain valid bundles and one negative vector for
each declared recall or mutation failure family. Cryptographic vectors add
signature, certificate, epoch, revocation, and trust-anchor checks. The
verification implementation is read-only and has no runtime importer, database,
network, signing, model, or policy authority.

\subsection{Production-derived conformance corpus}

A separate corpus projects production-shaped recall evidence into public test
objects and evaluates it with the production protocol owners and an independent
verifier. All \MutMemConformanceCount\ intended cases are observed, cover
\MutMemConformanceClassCount\ required classes, and agree on terminal verdict
and protocol reason. This establishes production/protocol conformance for the
released corpus. It does not establish retrieval quality or robustness against
unrepresented attacks. \evidence{V2-PRODUCTION-CONFORMANCE}

\subsection{Complexity and failure closure}

For $n=13+5r$ objects, membership construction uses maps keyed by singleton
kind and result ordinal, then performs one fixed-order traversal in $O(n)$ time.
The Merkle tree contains $n$ leaf hashes and $n-1$ internal hashes, hence
$O(n)$ hash evaluations. The current JavaScript reference uses recursive array
slices; accounting for those copies gives $O(n\log n)$ implementation time in
the worst case, $O(n)$ retained tree state, and $O(\log n)$ recursion depth,
with $r\le 200$. This bound is preferable to describing the hash-node count as
the complete runtime cost. The failure language is finite and versioned; new
predicates require a new vector and protocol root rather than an unrecorded
exception.

\section{Reproducibility architecture}

\subsection{Verification, installation, and reproduction are different}

The offline verification command checks retained public artifacts and executes
the independent Node/Python vector suites without opening AIMOS or a provider.
Evidence regeneration independently rebuilds publication tables, the attempt
ledger, disclosure boundary, and claim map. Neither command writes memory.

The installer is a separate product operation. The qualified run used Node
\MutMemInstallerNodeVersion, completed first boot and restart, registered the
required scheduler, retained \MutMemInstallerGuideCount\ Guide memories and
\MutMemInstallerExperimentalCount\ experimental memories, and capped the
ordinary enrolled agent at clearance \MutMemInstallerClearanceMaximum. This is
one clean-install qualification, not cross-platform coverage or a scientific
benchmark. \evidence{V2-CLEAN-INSTALLER}

Full reproduction is intentionally distinct from verification. It uses the
installer-selected ordinary agent, canonical signed SAVE and RECALL, fixed
dataset locks, fixed model roles, complete terminal evidence, and the integer
denominator invariant

\begin{equation}
n_{\mathrm{intended}}=n_{\mathrm{selected}}=n_{\mathrm{completed}}
=n_{\mathrm{evaluated}}, \qquad
n_{\mathrm{incomplete}}=n_{\mathrm{failed}}=0.
\label{eq:terminal-denominator}
\end{equation}

A mutable status flag is insufficient. The terminal binds the selected
immutable aggregate path and hash, required phase summaries, environment
identity, and the exact requested benchmark set. A stale predecessor cannot be
substituted for its completed immutable successor.

\subsection{Datasets, models, and environment}

The public package carries corpus manifests, source locks, target locks,
downloaders, and hashes. It does not redistribute LongMemEval, LoCoMo, Natural
Questions/BEIR, or PoisonedRAG text. Historical V1 model identities are retained
beside their results. Credentials, provider requests and responses, benchmark
text, memory identifiers, certificates, and machine paths are outside the
public evidence boundary.

Each new run must record operating system, architecture, CPU, memory, Node
binary, package and lock hashes, dependency versions, PostgreSQL and extensions,
concurrency, protocol configuration, model identities, native surfaces, and
start/capture times. This environment record is descriptive evidence, never
identity or configuration authority.

\section{Statistical contract}

For a binary outcome with $x$ successes in $n$ intended units, the estimate is
$\hat p=x/n$. The two-sided 95\% Wilson interval uses
$z=1.959963984540054$:

\begin{equation}
\frac{\hat p+z^2/(2n)\ \pm\ z
\sqrt{\hat p(1-\hat p)/n+z^2/(4n^2)}}{1+z^2/n}.
\label{eq:wilson}
\end{equation}

For paired binary outcomes with discordant counts $b$ and $c$, the exact
two-sided McNemar value is

\begin{equation}
p_{\mathrm{McN}}=\min\!\left(1,
2\sum_{k=0}^{\min(b,c)} {b+c\choose k}2^{-(b+c)}\right).
\label{eq:mcnemar}
\end{equation}

Within each declared family of $m$ contrasts, raw values are ordered
$p_{(1)}\le\cdots\le p_{(m)}$ and Holm adjustment is

\begin{equation}
\widetilde p_{(i)}=\max_{1\le j\le i}
\min\!\left(1,(m-j+1)p_{(j)}\right).
\label{eq:holm}
\end{equation}

Agreement uses $\kappa=(p_o-p_e)/(1-p_e)$ from public agreement counts and
positive marginals. Integers and hashes require exact equality; derived rates,
Wilson intervals, McNemar values, Holm adjustments, and $\kappa$ use numerical
tolerance $10^{-12}$. The historical Canary artifact stores Wilson bounds to
eight decimals; comparison to those source bounds uses $5\times10^{-9}$, while
the generated table retains full precision.

The independent evidence verifier reconstructs \MutMemStatisticalRateRows\
rate/Wilson rows and \MutMemStatisticalInferenceRows\ McNemar/Holm rows. V1
bootstrap intervals are not repeated in V2 tables because their private rows are
unavailable; retaining a historical point estimate is not the same as claiming
independent row-level regeneration.

\section{Evidence results}

\begin{table*}[t]
\centering
\caption{Current MutMem V2 reproducibility and conformance evidence. Counts are verification units, not benchmark accuracy.}
\label{tab:v2-evidence}
\small
\begin{tabular}{@{}>{\raggedright\arraybackslash}p{0.25\textwidth}>{\raggedright\arraybackslash}p{0.31\textwidth}>{\raggedright\arraybackslash}p{0.32\textwidth}@{}}
\toprule
Evidence family & Verified result & Interpretation \\
\midrule
Portable protocol & \MutMemProtocolSchemaCount\ schemas; \MutMemRecallStructuralVectorCount\ recall vectors; \MutMemMutationStructuralVectorCount\ mutation vectors; \MutMemFailureCodeCount\ recall failure codes & Versioned bytes, membership, predicates, and terminal vocabulary \\
Independent Node/Python verifier & \MutMemCrossLanguageTerminalCount/\MutMemCrossLanguageTerminalCount\ exact verdict and reason terminals; \MutMemVerifierExitPassed/\MutMemVerifierExitTotal\ exit criteria & Verifier portability, not empirical utility \\
Production conformance corpus & \MutMemConformanceCount/\MutMemConformanceCount\ cases across \MutMemConformanceClassCount\ required classes & Production/independent parity \\
Canary explicit-marker lane & marked \MutMemCanaryMarkedDetected/\MutMemCanaryMarkedTotal; clean flags \MutMemCanaryCleanFlagged/\MutMemCanaryCleanTotal & Explicit marker traversal only \\
Clean installer qualification & Node \MutMemInstallerNodeVersion; first boot, restart, and scheduler ready; \MutMemInstallerExperimentalCount\ experimental memories & One same-user installed system; no benchmark result \\
\bottomrule
\end{tabular}
\end{table*}

\subsection{Current V2 results}

Table~\ref{tab:v2-evidence} separates verification units from empirical
accuracy. Exact Node/Python parity and production conformance support a portable
verifier claim. They do not imply that another memory engine produces the same
utility results.

The Canary evidence contains \MutMemCanaryPopulation\ terminal units: all
\MutMemCanaryMarkedDetected/\MutMemCanaryMarkedTotal\ marked units were detected
(\MutMemCanaryMarkedPercent\%; Wilson lower
\MutMemCanaryMarkedWilsonLower\%), and
\MutMemCanaryCleanFlagged/\MutMemCanaryCleanTotal\ clean units were flagged
(\MutMemCanaryCleanPercent\%; Wilson upper
\MutMemCanaryCleanWilsonUpper\%). This supports only explicit marker traversal
across the declared transport cells. It is not evidence of arbitrary poison,
prompt-injection, or semantic social-engineering detection.
\evidence{V2-CANARY}

\subsection{Historical V1 evidence}

\begin{table*}[t]
\centering
\caption{Historical MutMem V1 results, reproduced from the self-hashed V1 aggregate. These are not current-source V2 reruns.}
\label{tab:v1-historical}
\small
\begin{tabular}{@{}>{\raggedright\arraybackslash}p{0.26\textwidth}>{\raggedright\arraybackslash}p{0.30\textwidth}>{\raggedright\arraybackslash}p{0.32\textwidth}@{}}
\toprule
Protocol and metric & Historical V1 result & Boundary \\
\midrule
LongMemEval, LLM-judged accuracy & \MutMemLongMemCorrect/\MutMemLongMemTotal\ (\MutMemLongMemPercent\%; Wilson 95\% CI \MutMemLongMemWilsonLower--\MutMemLongMemWilsonUpper\%) & GPT-5.4 generator; GPT-5.6 Terra judge \\
LoCoMo, LLM-judged accuracy & \MutMemLocomoCorrect/\MutMemLocomoTotal\ (\MutMemLocomoPercent\%; Wilson 95\% CI \MutMemLocomoWilsonLower--\MutMemLocomoWilsonUpper\%) & Distinct from token F1 \\
LoCoMo, category-aware token F1 & \MutMemLocomoFOne & Bootstrap interval omitted in V2: private rows unavailable \\
PoisonedRAG adaptation, induced ASR & \MutMemPoisonInducedCount/\MutMemPoisonInducedTotal\ (\MutMemPoisonInducedPercent\%) & Baseline-aware effect: clean 2/100; attacked 3/100 \\
PoisonedRAG adaptation, poison retrieval@5 & \MutMemPoisonRetrievalCount/\MutMemPoisonRetrievalTotal\ (Wilson upper \MutMemPoisonRetrievalWilsonUpper\%) & Retention plus retrieval isolation, not deletion \\
Mutation integrity & \MutMemMutationAuthorizationPassed/\MutMemMutationAuthorizationTotal\ authorization; \MutMemMutationTamperPassed/\MutMemMutationTamperTotal\ tamper cases & \MutMemMutationTransitionCount\ measured native transitions \\
Mutation transaction cost & median \MutMemMutationMedianMs\ ms; p95 \MutMemMutationPNinetyFiveMs\ ms; \MutMemMutationMeanBytes\ logical bytes/transition & Single historical machine and full transaction \\
Blinded system-author agreement & \MutMemHumanAgreementCount/\MutMemHumanAgreementTotal\ (\MutMemHumanAgreementPercent\%; $\kappa=\MutMemHumanKappa$) & Not independent human validation \\
\bottomrule
\end{tabular}
\end{table*}

Table~\ref{tab:v1-historical} is intentionally labeled historical. LongMemEval
and judged LoCoMo used the historical canonical-blind protocol; the LoCoMo token
F1 result used a separate deterministic scorer and must not be averaged with
judged accuracy. \evidence{V1-LONGMEMEVAL-UTILITY}
\evidence{V1-LOCOMO-JUDGED}\evidence{V1-LOCOMO-F1}

The historical N=100 PoisonedRAG result is an adaptation: target fixture and
attacker passages were locked, but corpus scope, retriever, and answer model
differed from the upstream full-corpus experiment \cite{poisonedrag}. All poison
passages were retained. Retrieval isolation and signed epistemic labels are not
save-time rejection, deletion, or proof that content is false. The observed
clean target-answer leakage was \MutMemPoisonCleanAsrCount/\MutMemPoisonCleanAsrTotal,
while the attacked target-match rate was
\MutMemPoisonAttackedAsrCount/\MutMemPoisonAttackedAsrTotal. The baseline-aware
attack-attributable result was therefore an induced ASR of
\MutMemPoisonInducedCount/\MutMemPoisonInducedTotal\
(\MutMemPoisonInducedPercent\%), which is the effect reported in
Table~\ref{tab:v1-historical}. The observed
clean and attacked answer accuracies were \MutMemPoisonCleanAccuracyPercent\%
and \MutMemPoisonAttackedAccuracyPercent\%, respectively; all
\MutMemPoisonLabelCount/\MutMemPoisonLabelTotal\ poison memories received an
adverse signed projection, and \MutMemCleanAdverseCount/\MutMemCleanAdverseTotal
clean memories did as well (\MutMemCleanAdversePercent\%).
\evidence{V1-POISONEDRAG}

\begin{table}[t]
\centering
\caption{Historical V1 fixed-corpus PoisonedRAG ablation. Each arm contains 100 paired targets.}
\label{tab:v1-ablation}
\small
\begin{tabular}{@{}lrrr@{}}
\toprule
Arm & poison retrieval@5 & clean accuracy & attacked accuracy \\
\midrule
A0 & \MutMemAblationAZeroRetrievalCount/\MutMemAblationAZeroRetrievalTotal & \MutMemAblationAZeroCleanPercent\% & \MutMemAblationAZeroAttackedPercent\% \\
A1 & \MutMemAblationAOneRetrievalCount/\MutMemAblationAOneRetrievalTotal & \MutMemAblationAOneCleanPercent\% & \MutMemAblationAOneAttackedPercent\% \\
A2 & \MutMemAblationATwoRetrievalCount/\MutMemAblationATwoRetrievalTotal & \MutMemAblationATwoCleanPercent\% & \MutMemAblationATwoAttackedPercent\% \\
A3 & \MutMemAblationAThreeRetrievalCount/\MutMemAblationAThreeRetrievalTotal & \MutMemAblationAThreeCleanPercent\% & \MutMemAblationAThreeAttackedPercent\% \\
\bottomrule
\end{tabular}
\end{table}

The historical fixed-corpus ablation in Table~\ref{tab:v1-ablation} attributes
the measured selection change to already-produced signed stored labels. A2
added query-local detection but made no additional selection change after A1;
active-context withholding was not exercised. The experiment therefore does not
establish how either mechanism behaves on unseen attacks.
\evidence{V1-EPISTEMIC-ABLATION}

The historical mutation measurements cover one complete native transaction,
including signatures, provenance, projection, and live-weight update. They are
not isolated cryptographic overhead and should not be generalized across
machines. \evidence{V1-MUTATION-LATENCY-STORAGE}

The historical agreement audit was blinded to arm and judge verdict but was
performed by the system author. It is an agreement diagnostic, not independent
human validation. \evidence{V1-HUMAN-AGREEMENT}

\section{Attempt accounting and disclosure}

The public attempt ledger preserves intended population, terminal population,
terminal failures, retries when known, exclusions, amendments, and reused
artifacts. Two failed installed-service attempts remain failures with zero result
authority. A later one-question diagnostic remains diagnostic only. None appears
in a result table, abstract, or comparison. For promoted numerical evidence,
intended and terminal populations are equal, with zero terminal failures and
zero exclusions. Unknown historical retry details are explicitly unknown, not
zero.

The public/private manifest withholds private V1 rows, provider payloads,
restricted dataset text, credentials and certificates, live memories, and
identity-bearing installer receipts. It exposes sanitized aggregates, semantic
roots, source locks, and non-identifying readiness projections instead. The
SABER-inspired operational figures are omitted because their exact private
artifact is not in verified custody; no value from that family appears in this
paper. \evidence{SABER-OPERATIONAL-NUMBERS}

\section{Security argument and non-goals}

\textbf{Commitment substitution.} Under collision resistance, changing a
canonical object field changes Equation~\ref{eq:object}, the object root, and
Equation~\ref{eq:bundle}, except with negligible collision probability.
Omitting or duplicating a mandatory object fails Equation~\ref{eq:cardinality}
before signature verification.

\textbf{Authority substitution.} A bundle does not choose its own trusted
master. Actor and Housekeeper certificate chains, validity intervals,
revocation observations, grant scope, signed request, and terminal event are
checked against the externally selected trust anchor. Under Ed25519
unforgeability, altering a signed commitment without the corresponding private
key fails signature verification.

\textbf{Ordering and terminal fidelity.} Per-result ordinals and fixed kind
order enter the Merkle root. The terminal event binds the request, authority,
root, evidence list, return projection, and result count. Reordering a valid set
or replacing the terminal projection is detectable.

These arguments establish tamper evidence under the assumptions. They do not
prevent a storage administrator from destroying evidence, prove that signed
content is true, or guarantee that every harmful input is detected. Availability
and semantic validation require separate mechanisms and evidence.

\section{Limitations}

\begin{enumerate}
  \item \textbf{No current V2 utility rerun.} Utility, mutation performance,
  PoisonedRAG, ablation, and human agreement remain historical V1 results. V2
  does not relabel them as results of the current source.
  \item \textbf{No independent empirical replication.} The verifier
  implementations are independent, but the empirical experiments have not been
  repeated by an outside team.
  \item \textbf{Private V1 rows are unavailable.} Public counts, Wilson
  intervals, McNemar/Holm contrasts, and agreement marginals reconstruct. V1
  fixed-seed bootstrap intervals do not independently reconstruct and are
  omitted from V2 tables.
  \item \textbf{Canary scope is marker-specific.} Marker traversal is not a
  general detector for poisoning, prompt injection, or false content.
  \item \textbf{PoisonedRAG is adapted and post-calibration.} It does not match
  the upstream full corpus and does not estimate unseen-attack generalization.
  \item \textbf{SABER figures are omitted.} Artifact custody is insufficient
  for numerical publication.
  \item \textbf{Installation coverage is bounded.} One macOS installation on
  Node \MutMemInstallerNodeVersion\ is qualified. Other declared runtimes still
  require release CI and do not inherit this result automatically.
  \item \textbf{External anchors are necessary.} Replacing all evidence and all
  independent trust anchors is outside the threat model.
  \item \textbf{Archival identifier pending.} The immutable source manifest,
  release commit, tag, and GitHub Release are mutually bound. The arXiv
  identifier cannot be recorded until submission acceptance.
\end{enumerate}

\section{Ethics and responsible release}

The poisoning materials contain attacker-crafted misinformation and restricted
dataset content. The release distributes hashes, manifests, aggregate outcomes,
and acquisition instructions rather than benchmark passages or provider
payloads. Reproduction must use a fresh installed brain and must not ingest
benchmark material into a user's ordinary persistent memory. A detected item is
retained with signed epistemic evidence; classification is reversible and is
not presented as factual adjudication.

The verifier's public failure reasons are designed for audit, but operational
receipts may contain identity and storage metadata. Such receipts remain outside
the public evidence allowlist. Public evidence contains no credential, private
key, certificate, live memory content, or absolute machine path.

\section{Related work}

Tamper-evident logging uses cryptographic commitments to make history
substitution detectable \cite{crosbywallach}. Certificate Transparency defines
the domain-separated Merkle construction used here for ordered evidence
\cite{rfc6962}. MutMem combines established SHA-256, Ed25519 \cite{rfc8032},
canonical JSON \cite{rfc8785}, and length-framed protocol encodings
\cite{bonehshoup}; none of those primitives is claimed as novel.

LongMemEval and LoCoMo measure long-horizon memory utility
\cite{longmemeval,locomo}. PoisonedRAG studies knowledge-corruption attacks
against retrieval-augmented generation \cite{poisonedrag}. Those works motivate
evaluation but do not by themselves provide a portable authorization and
evidence profile for post-deployment memory mutation. MutMem V1 introduced the
retained mutation construction and historical system evaluation
\cite{mutmemv1}. V2's narrower contribution is to make its evidence boundaries,
canonical bytes, terminal reasons, independent verification, installation, and
publication derivation explicit.

\section{Conclusion}

MutMem V2 turns an implementation-bound integrity claim into a versioned,
externally anchored verification protocol. Complete recall and mutation bundles
bind canonical objects, cross-object authority, result identity, ordered
disclosure, and terminal events. Independent Node and Python implementations
agree on the full released terminal corpus, production-derived cases agree with
the independent verifier, and one clean installer reaches restart-stable
readiness. Publication tables, attempts, omissions, and limitations regenerate
from a single evidence root.

The result is a stronger reproducibility claim, not a broader security claim.
Historical V1 empirical findings remain historical. Canary evidence remains
marker-specific. SABER numerical evidence remains omitted. Integrity and
authorization remain distinct from truth, and verifier independence remains
distinct from independent empirical replication. The public release identity
is mutually bound; the archival identifier is recorded after acceptance.

\appendix
\section{Complete claim-to-evidence disposition}

\scriptsize
\begin{longtable}{@{}>{\raggedright\arraybackslash}p{0.16\linewidth}>{\raggedright\arraybackslash}p{0.12\linewidth}>{\raggedright\arraybackslash}p{0.14\linewidth}>{\raggedright\arraybackslash}p{0.42\linewidth}@{}}
\caption{Complete manuscript claim-to-evidence disposition.}\label{tab:claim-map}\\
\toprule
Claim identifier & Disposition & Denominator & Limitation \\
\midrule
\endfirsthead
\toprule
Claim identifier & Disposition & Denominator & Limitation \\
\midrule
\endhead
V1-\allowbreak{}LONGMEMEVAL-\allowbreak{}UTILITY & HISTORICAL V1 & 459/500 & Historical V1 result; model-judged and not a current-source V2 benchmark. \\
V1-\allowbreak{}LOCOMO-\allowbreak{}JUDGED & HISTORICAL V1 & 1472/1986 & Historical V1 judged accuracy; never merged with token F1. \\
V1-\allowbreak{}LOCOMO-\allowbreak{}F1 & HISTORICAL V1 & 1986 & Point estimate remains in the V1 aggregate; bootstrap interval omitted because private rows are unavailable. \\
V1-\allowbreak{}POISONEDRAG & HISTORICAL V1 & 100 targets & Adapted bounded N=100 protocol; no universal or unseen-attack claim. \\
V1-\allowbreak{}EPISTEMIC-\allowbreak{}ABLATION & HISTORICAL V1 & 100 paired targets per arm & Fixed-corpus causal evidence; query-local contribution was null and withholding was not exercised. \\
V1-\allowbreak{}MUTATION-\allowbreak{}INTEGRITY & HISTORICAL V1 & 20 transitions & Historical retained-memory transition evidence under the V1 threat model. \\
V1-\allowbreak{}MUTATION-\allowbreak{}LATENCY-\allowbreak{}STORAGE & HISTORICAL V1 & 20 transitions & Single-machine descriptive values; row-level inputs are unavailable for independent recomputation. \\
V1-\allowbreak{}HUMAN-\allowbreak{}AGREEMENT & HISTORICAL V1 & 200 answers & System-author diagnostic, not independent human validation; bootstrap interval omitted. \\
V2-\allowbreak{}PORTABLE-\allowbreak{}PROTOCOL & CURRENT V2 & 39+15 vectors & Portable verification protocol, not another memory engine or producer. \\
V2-\allowbreak{}INDEPENDENT-\allowbreak{}VERIFIER & CURRENT V2 & 72 terminal verdicts & Cross-language verdict/reason parity proves verifier portability, not empirical utility. \\
V2-\allowbreak{}PRODUCTION-\allowbreak{}CONFORMANCE & CURRENT V2 & 42/42 & Conformance corpus, not benchmark efficacy. \\
V2-\allowbreak{}CANARY & CURRENT V2 & 120/120 & Explicit marker traversal only; not arbitrary poison, prompt-injection, or semantic social-engineering detection. \\
V2-\allowbreak{}CLEAN-\allowbreak{}INSTALLER & CURRENT V2 & one qualified installation and restart & Reproducibility evidence only; not benchmark evidence. \\
SABER-\allowbreak{}OPERATIONAL-\allowbreak{}NUMBERS & OMITTED & No numerical claim & Exact private artifact is not in verified custody; no numerical V2 claim is permitted. \\
CONTENT-\allowbreak{}TRUTH & PROHIBITED & No numerical claim & Integrity, provenance, authorization, and agreement do not establish semantic truth. \\
UNIVERSAL-\allowbreak{}DEFENSE & PROHIBITED & No numerical claim & Fixed marker and N=100 evidence do not establish universal robustness. \\
INDEPENDENT-\allowbreak{}REPLICATION & PROHIBITED & No numerical claim & No outside team independently reproduced the system or empirical results. \\
\bottomrule
\end{longtable}

\section{Verifier and regeneration commands}

\begin{verbatim}
npm run verify
npm run evidence:regenerate
npm run verify
npm run paper:verify
\end{verbatim}

The first, third, and fourth commands are read-only. The second command
deterministically rewrites only the public evidence projections and fails its
subsequent byte check if output changes. No command in this appendix runs a
benchmark.

\end{document}